# ECONOMIC AMPLIFIER – A NEW ECONOPHYSICS MODEL

Ion SPÂNULESCU* and Anca GHEORGHIU

Hyperion University of Bucharest, 169 Calea Călărașilor, 030629, Bucharest, Romania

*Keywords:* econophysics models, electronic amplifier, economic amplifier, $\beta$ -economic amplification coefficient

**Abstract.** Most of the econometric and econophysics models have been borrowed from the statistical physics, and as a cosequence, a new interdisciplinary science called econophysics has emerged. In this paper we planned to extend the analogy between different economic processes or phenomena and processes/phenomena from different fields of physics, other than statistical physics. On the basis of the economic development process and amplification phenomenon analogy, a new econophysics model, named "economic amplifier", on the electronic amplification principle from applied physics was proposed und largely analyzed.

## 1. Introduction

During last decade, on the basis of many studies and published papers and books, a new interdisciplinary science has appeared named econophysics, which uses models taken especially from statistical physics to describe some economic phenomena and processes.

Most econophysics approaches, models and papers that have been written so far refer to the economic processes including systems with a large number of elements (such as financial or banking markets, stock markets, incomes, production or product's sales, individual incomes etc.) where statistical physics methods, Boltzmann, Gibbs and some other statistical distribution types are mainly applied [1÷19].

Reducing econophysics only to of statistical physics applications in economy, especially in the stock markets analysis, seems to be very restrictive. It is desirable to investigate other fields of physics, and economy too, in which processes similarity inspire and even facilitate to adopt new econophysics models using – where it is possible – some other physics fields too.

This paper proposes an extension of analogy between different economic phenomena and processes and phenomena or processes form different fields of physics, other than statistical physics. In this way, taking in consideration the importance of economic development model and production amplification, a model based on amplification concept from electronic physics, named **economic amplifier** was proposed and discussed. The model implies an electronic amplifier realized with active electronic devices, the most by used being the device named transistor, which represents itself an excellent amplifier, of both current and electrical power [20, 21].

The transistor and the electronic amplifier made of transistors, presents the features of an exact model, a determinist one. Thus, if the transistor is used for economic modeling, it guarantees the same accuracy and validity to the economic models to which the economic amplifier model is correlated and assimilated.

## 2. Electronic Amplifier

Any electronic amplifier has an active element which is either bipolar transistor (with *p-n* junctions) or field effect transistor (MOS type).

In this section we are going to present some parameters and characteristics of bipolar transistor with *p-n* junctions and the MOSFET structure and characteristics.

### 2.1. The bipolar transistors

As it can see in figure 1*a*, the bipolar transistor is made of two semiconductor *p-n* junctions, first junction (emitter junction) being forward biased with $V_{BE}$ voltage, and the second one (collector junction) being reverse biased with $V_{CB}$ voltage, with

$$|V_{CB}| \gg |V_{BE}|. \tag{1}$$

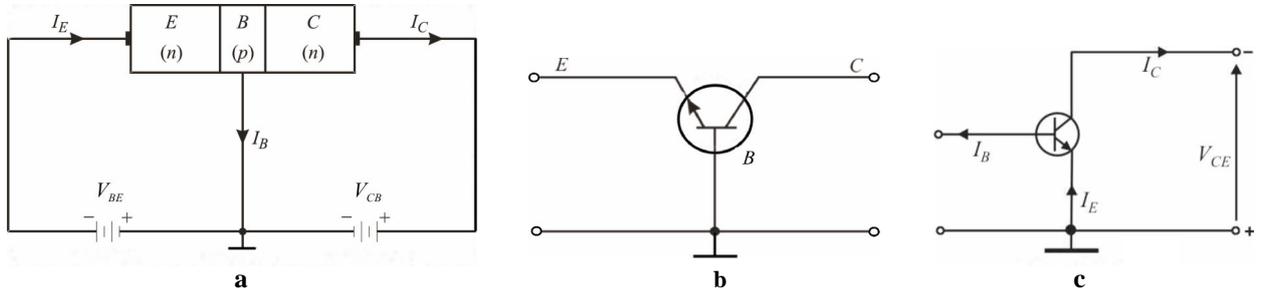

**FIG. 1** *a*) The structure and bias for the *n-p-n* bipolar transistor; By proper biasing of the junction, the transistor will be crossed by emitter, base and collector currents ($I_E$, $I_B$ and $I_C$) which are in the following relation

$$I_E = I_B + I_C \tag{2}$$

in which $I_E$ represent the current due to injection of charge carriers from emitter region to base region where they are transferred into collector circuit as collector current $I_C$ under the action of electric field of collector junction (Fig. 1*a*); *b*) the symbol for the *n-p-n* bipolar transistor; *c*) The *n-p-n* transistor common emither (CE) configuration.

Due to inverse currents of minority carriers from collector and emitter regions and recombination current of charge carriers in the base region, the base current $I_B$ is much smaller than $I_E$ or $I_C$

$$\begin{aligned} I_B &\ll I_C \\ I_E &\approx I_C. \end{aligned} \tag{3}$$

Considering the transistor as a quadruple there are three ways to connect it in circuit, one of electrodes being common to both input and output circuits: CB configuration (with common base), CE configuration (with common emitter) and common collector configuration (CC). In amplification circuits the common emitter configuration is mostly used because it provides a highest power amplification (gain) (Fig. 1*c*).

For transistor analysis we shall use the Ebers-Moll equations for static regime (with the biasing voltages only) [21, 22]

$$I_E = I_{ES}\left[\exp\left(\frac{eV_{BE}}{kT}\right) - 1\right] - \alpha_I I_{CS}\left[\exp\left(\frac{eV_{CB}}{kT}\right) - 1\right] \tag{4}$$

$$I_C = \alpha_N I_{ES}\left[\exp\left(\frac{eV_{BE}}{kT}\right) - 1\right] - I_{CS}\left[\exp\left(\frac{eV_{CB}}{KT}\right) - 1\right] \tag{5}$$

in which

$$\alpha_N = \frac{I_C}{I_E} \leq 1 \tag{6}$$

is the "normal" current amplification coefficient for transistor in BC configuration in static regime with properly biasing of the junctions;

$\alpha_1$ is the reverse current amplification coefficient when the collector junction is forward biased ($V_{CB} > 0$), and the emittor junction is reverse biased; because the collector is very little doped, usually

$$\alpha_1 << \alpha_N. \tag{7}$$

The current amplification coefficient for transistor in CE configuration represent the ratio between output current $I_{out}$, that is collector current $I_C$, and input current $I_{in}$ that is $I_B$

$$\beta = \frac{I_{out}}{I_{in}} = \frac{I_C}{I_B}. \tag{8}$$

For big values of negative biasing voltage applied on the collector junction, it can be written

$$|-V_{CB}| >> KT/e \tag{9}$$

and as a result, equations (4) and (5) become:

$$I_E = I_{ES}\left[\exp\left(\frac{eV_{BE}}{kT}\right) - 1\right] \tag{10}$$

$$I_C = \alpha_N I_{ES}\left[\exp\left(\frac{eV_{BE}}{kT}\right) - 1\right]. \tag{11}$$

From Eq.(1), $I_B$ can be teken as

$$I_B = I_{ES}(1 - \alpha_N)\left[\exp\left(\frac{eV_{BE}}{kT}\right) - 1\right]. \tag{12}$$

Using the Eqs (11) and (12) the current amplification coefficient defined by Eq.(8) becomes

$$\beta = \frac{I_C}{I_B} = \frac{\alpha_N I_{ES}\left[\exp\left(\frac{eV_{BE}}{kT} - 1\right)\right]}{(1 - \alpha_N)I_{ES}\left[\exp\left(\frac{eV_{BE}}{kT} - 1\right)\right]} = \frac{\alpha_N}{1 - \alpha_N}. \tag{13}$$

According to Eq.(6), $\alpha_N < 1$, so it results that $\beta$ from Eq.(13) is grater than unit

$$\beta = \frac{\alpha_N}{1 - \alpha_N} >> 1. \tag{14}$$

Usually, for the good bipolar transistors $\beta$ varied between 100 and 1000 or more. From Eq.(14) it can be noticed that transistors in the CE configuration is a very good current amplifier.

Generally, the transistor, like any other amplifier, is used to amplify variabile signals, $v_{in}$, all these being much smaller then biasing voltages in continuous current (c.c.)

$$|v_{in}| << V_{BE}. \tag{15}$$

In this case we have the dynamic regime of transistor functioning to small signals, when voltages and/or variable currents can be considered as little variations of the biasing currents or voltages

$$v_{in} = v_{BE} = \Delta V_{BE}; \quad i_C = \Delta I_C \text{ etc.} \qquad (16)$$

In dynamic regime for variable voltages and currents small letters are used for noticing magnitudes, meaning $i_b, v_{in}, i_c, v_o$ etc.

The following parameters of bipolar transistors functioning, in the dynamic regime, can be defined [21, 22]:

1) input resistance $r_{in}$, defined through the ratio between input voltage variation an input current variation

$$r_{in} = \frac{\partial v_{BE}}{\partial i_b} = \frac{\partial v_{in}}{\partial i_b}; \qquad (17)$$

2) output conductance is given by

$$G_{out} = \frac{1}{r_{out}} = \frac{\partial i_c}{\partial v_{bc}}; \qquad (18)$$

3) slope or transconductance for bipolar transistors are given by [21, 22]

$$S = g_m = \frac{\partial i_c}{\partial v_{be}}. \qquad (19)$$

These parameters can be also defined for static regime when only the biasing voltage are applied. For finite variations of currents and voltage, in the static regime the above relations can be written as:

$$r_{in} = \frac{\Delta V_{BE}}{\Delta I_B}; \quad \frac{1}{r_{out}} = \frac{\Delta I_C}{\Delta V_{BC}}$$

and

$$S = \frac{\Delta I_C}{\Delta V_{BE}}. \qquad (20)$$

As it can be seen in Eq.(20) the slope $S$ shows how much collector current (output current) increases when the input voltage ($V_{BE}$) increases by a unit.

### 2.2. Unipolar Metal-Oxide-Semiconductor (MOS) transistors

The unipolar transistors also contain two *p-n* junctions but they are functioning with only **one** charge carriers type, the majority ones (where their name came from) on the basis of field effect, being named Field Effect Transistors (FET) also.

There are several types of these transistors but the most often used in current applications is Metal-Oxide-Semiconductor (MOS) transistor where control electrode named Gate is isolated from the device through an oxide or dielectric thin film (Fig. 2*a*).

The analog of emitter from bipolar transistor is the Source (S) where charge carriers are coming from, and the analog of collector is the Drain (D) that is collecting the carriers (Fig. 2).

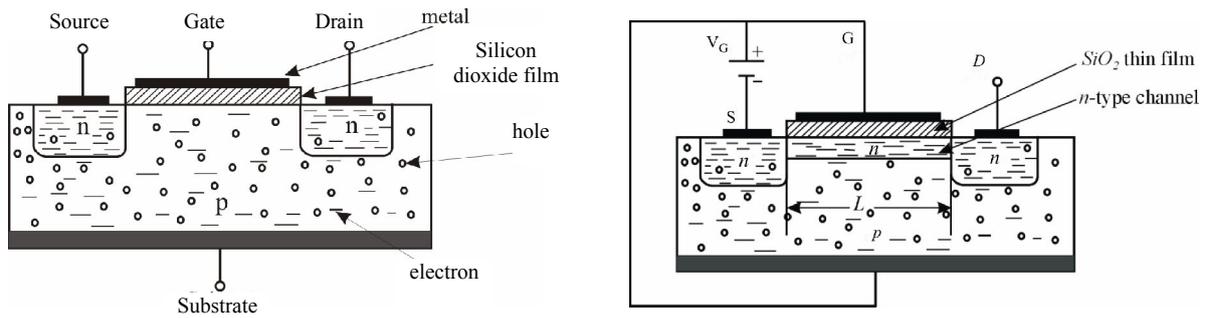

**FIG. 2** *a)* FET structure of *n* channel-MOS type; *b)* The conduction channel induced in the MOSFET structure.

Initially, the source and drain regions are completely isolated so there is no current flowing in the MOS structure from figure 2*a*, between source and drain. If a positive voltage $V_{GS}$ is applied between gate and source there will appear a conduction channel of *n*-type between source and drain where the electrons there the majority carriers will be able to flow. This transistor is named MOSFET with *n*-channel or NMOSFET.

The channel that has been formed is one that has been induced due to gate positive potential action that will reject the holes from the region placed exactly under oxide to semiconductor depth of *p*-type substrate. The positive gate voltage will attract the mobile electrons from the *p*-substrate in this region which being in large number under oxide will create a *n*-region that represents the conduction channel between source and drain (Fig. 2*b*).

After channel creation between drain and source, if a positive voltage $V_{DS} > 0$ is applied, the electrons injected into the source (by biase voltage) will flow to drain through channel, and therefore a drain current $I_D$ will result (Fig. 3). By varying input voltage $V_{GS}$ (which includes a variable voltage $v_{in}$ that must by amplified), the thickness and electrical resistance of induced *n*-channel will vary (Fig. 3) and, as a consequence, the drain current $I_D$ will vary too. In this way, an amplified output voltage $U_{out}$ can be taken from load resistor $R_L$ (Fig. 3):

$$U_{out} = I_D R_L. \qquad (21)$$

From figure 3 it can be seen that as result of applying voltage $V_{DS} > 0$, the *n*-conduction channel is narrowing to collector region limiting $I_D$ current reaching the saturation value $I_{D\,sat}$ (Fig. 3) [25,26].

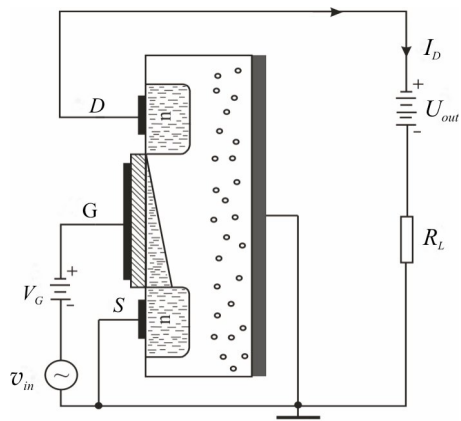

**FIG. 3** MOSFET in the amplification regime $(I_D \neq 0)$.

For the MOS transistor the slope $S$ or transconductance $g_m$ is defined as [21-23]

$$g_m = S = \frac{\partial I_D}{\partial V_{GS}}. \tag{22}$$

PMOSFET (*p*-MOS Field Effect Transistor) with *p*-type induced channel is functioning in the same way where source and drain regions of *p*-type are diffused in a substrate of *n*-type and bias voltages are opposite to those of NMOSFET, so $V_{GS} < 0$ and $V_{DS} < 0$. In both cases, MOS transistors are functioning with only one type of charge carriers: electrons or holes, so they are unipolar transistors.

### 2.3. Electronic amplifier with bipolar transistors

Figure 4 represent a practical circuit of an amplifier with bipolar transistor. To avoid using a supplementary source for $V_{BE}$ voltage, a voltage divider $R_{B_1}, R_{B_2}$ for proper biasing of base (input circuit) in used. The amplifier has only one amplification stage because it uses only one transistor (active element) for amplification. To increase the amplification, the amplifiers with two or more transistors, having two or more amplification stages are used.

If a small variable signal $\mathbf{v}_{in}$ is applied to amplifier input (Fig. 4) an amplified signal is taken from load resistor $R_L$ as output voltage

$$\mathbf{v}_0 = i_c R_L = U_{out}. \tag{23}$$

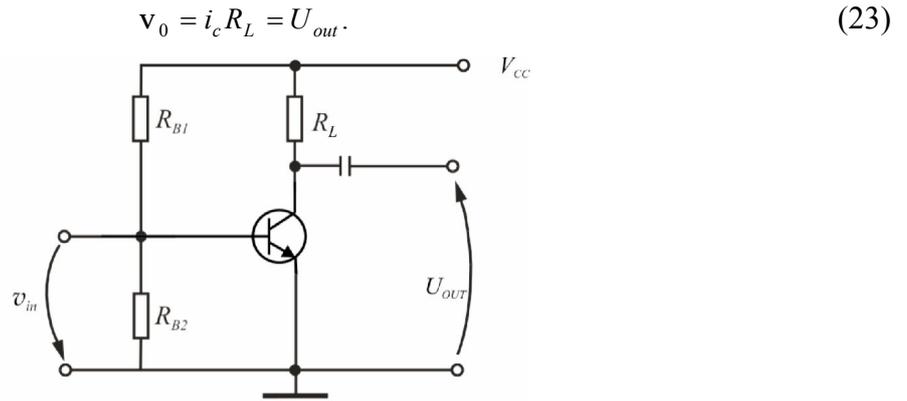

**FIG. 4** Amplifier circuit with a bipolar transistor in CE configuration.

The current amplification coefficient is given by Eq.(8) that is practically the same with current amplification coefficient for the variable signals

$$\beta = \frac{i_c}{i_b}. \tag{24}$$

Also, an amplified power is put out on the load resistor $R_L$ (Fig. 4):

$$P_{out} = U_{out} i_c = i_c R_L i_c = R_L i_c^2 \tag{25}$$

these obtaining a power gain of the signal applied to amplifier input.

## 3. Economic Amplifier

In this paper a model based on transistor-effect from physics, meaning **amplification phenomenon** that can be realized using transistors, was proposed. Such a model can be used for modeling different economic structures or processes such as production or investment fields, steady capital, founds or financial – banking fields, stocks etc.

Transistor-effect of amplification of transistor device is completely verified in practical activities. Modeling physics phenomena on the electronic level for bipolar and unipolar transistors theory is validated through all technological and socio-economical developments of nowadays society, this little device being the base brick of all electronic apparatus and equipments in any field of activity. The same overwhelming confirmation of transistor model validity (which is multiplied in hundreds and thousand millions of components into microprocessors or some other integrated systems) gives to the transistor the leader status of all contemporary discoveris and practical applications in physics and techniques. That's why the transistor deserves the whole attention of researchers that are studying models of different human activity phenomena and processes, especially economics, finances, management etc. The almost universal validity of amplifying phenomenon, in particular by transistor-effect, guarantees the validity of economic model action on the basis of electronic amplifier with transistors, that we named "economic amplifier model".

First example of similitude and modeling with transistor model is represented by assimilation of charge carriers in transistors with the number of products realized inside a section or a factory etc. by an economic unit. As charge carriers current flows through transistor from input to output (Fig. 1*a*), the products flow from input (where they represent parts, raw materials etc.) to the output of a section or a manufacture and then all this stuff is delivered as finite products.

The role of load resistor $R_L$ from output circuit (Fig. 4) is taken by storage or transportation systems (conveyors, containers, trains etc.) which deliver products to consumers in the same way in which charge carriers are collected from load resistor as current intensity $I_C$.

The expression $I_C R_L$ represents the output voltage being assimilated with technical platform – production ensemble from workshop, manufacture or factory etc. where equipments, labour force (human factor) and other production expenses are included.

The **charges** carried by charge carriers are assimilated with **values** that final products are carrying that represent "valor carriers".

The economic amplifier can be interpreted as an Input-Output model like one stage transistor electronic amplifier. So, economic amplifier model can be analyzed using Eq.(8) for current amplification coefficient β, of bipolar transistor

$$\beta = \frac{I_{out}}{I_{in}}. \tag{8'}$$

In this way if we consider that at the input circuit, between base and emitter a signal assimilated with investments (capital or in nature) introduced into factory (economic unit) is applied and at the output as current $I_{out}$ we consider the total amount of products or total income obtained, we can define so called $\beta_{economic}$ coefficient. This coefficient can be defined for all quantity of products $\beta_p$ as

$$\beta_{p\ economic} = \frac{\text{Total finished products}}{\text{Investments, raw materials and equipments}}. \tag{26}$$

A more important parameter can be introduced being the ratio of the value efficiency determined by the total amount of incomes (business value) shared over the investment amount, marked by $\beta_v$

$$\beta_{v\ economic} = \frac{\text{Total incomes}}{\text{Investments + expences}}. \tag{27}$$

The unipolar transistors of MOS type are most suitable econophysics model for describing banks and other financial institutions opperating with **only one** economic category: money, because the unipolar transistors operate with only one type of charge carriers [20÷23]. Here, similar to bipolar transistor, we work with financial investments as money on the input allowing to obtain a big total output income after a period of time (a year, a month etc.) so, a biger than unity gain is obtained. The growth (amplification) of the invested capital under the action of interests, commissions etc. is equivalent to the amplification concept from applied physics (electronics). Otherwise, the **gain** obtained from banks, stocks etc. is similar to the **gain** obtained from transistor amplifiers where we have a gain for bank (equivalent with the gain for current, β given by Eq.(8))

$$\beta_{bank} = \frac{\text{Output values (money, interests etc.)}}{\text{Total values (Initial capital + Amount obtained + Given Interests)}}. \quad (28)$$

Obviously, the amplification factor $\beta_{bank}$ is calculated for a period of standard time (for example: a year, a month, a week etc.) that is, over the time, of the financial operations execution.

The bankruptcy of an economic agent (firm, company) is modeling by the transistor breakdown or circuit failure. Circuit damage analysis could inspire solutions for bankruptcy avoidance. Transistor failure (bipolar or unipolar) could come from exploitation errors (improper voltages), supercharges, design errors, or because of semiconductors material quality, that is, for the reasons associated to the improper structure and design of the material, device (transistor) or circuit (amplifier). Similar these types of errors are also met because the managerial drawbacks, lower staff qualification, lack of communication, improper average age for a specific production stage of the company etc.

The modeling using more transistors and more amplifying stages can be used for the bigger economic systems such as big factories or corporations etc. So, one stage amplifier (using one transistor) is modeling one section activity. The multi-stage amplifiers having a big gain can simulate a factory with many sections. The bigger amplifier with many stages including the output power one are used to model a corporation or a big trust having a very large gain (amplification coefficient) manufacturing a large amount of products.

## 4. The correlation of "economic amplifier" model with other econometric models

Analyzing the characteristics and principal parameters of transistors in the amplification regime, described in 2.1 section, one can observe a good correspondence between the economic amplifier model and other projective models from the investments field described elsewhere [24,25].

### 4.1. The correlation with Harrod's investments model

Harrod considers that between operating assets and incomes, in the conditions of a neutral technical progress and unchanged interest installments, there exists a constant ratio, measured through the capital coefficient b, given by relation [24÷26]

$$b = \frac{\text{Investments (operating assets)}}{\text{Incomes}}. \quad (29)$$

Comparing Eq.(29) with β-economic expression given by Eq.(27)

$$\beta_{economic} = \frac{\text{Total incomes}}{\text{Total investments}}, \quad (27)$$

consequent upon this

$$\beta_{economic} = b^{-1} \tag{30}$$

meaning that economic amplifier model is complementary to Harrod's model.

### 4.2. The corespondance with Domar model (the investments productivity)

The Domar model is defined by the following equation [24÷26]

$$\sigma = \frac{\Delta Q}{I}, \tag{31}$$

in which:
- $\sigma$ – represents the investments productivity;
- $\Delta Q$ – production efficiency subsequent to be realized;
- $I$ – total investments.

Comparing with the economic amplifier, the production efficiency $\Delta Q$ can be identified with the variation of transistor output current, meaning – as it can be seen from Eq.(26) – that $\Delta Q$ corresponds to total incomes efficiency in the economic amplifier model.

Since for both models, the denomitor contains the investments volume (*I*, in the Domar model), the Eq.(31) in the case of the Domar model is the some as Eq(26) or Eq.(26) (after case) for β<sub>economic</sub>, in the economic amplifier model meaning that

$$\sigma = \beta_{economic}. \tag{32}$$

Therefore, according to Eq.(32), β-economic from the economic amplifier model is identified to the investments productivity, σ, from Domar's model.

### 4.3. The correlation of economic amplifier model with Cobb-Douglas production function

The Cobb-Douglas model for production function is given by relation [24÷26]

$$Q = gL^{\lambda}K^{\mu} \tag{33}$$

where $Q$ represent production, $K$ – operating capital, $L$ – labour force, $g$ – proportionality coefficient, $\lambda$ and $\mu$ – the elasticity coefficients of labour and operating capital respectivlly [24÷26].

Cobb-Douglas function can be modeled using the electrical power model produced on a load resistance $R_L$ at the output of an electronic amplifier (Fig. 4) whence it is outside transferred

$$P_{out} = UI_C. \tag{34}$$

By identification of the variables from Eq.(33) (without $\lambda$ and $\mu$ powers) with the variables from the Eq.(34) the production $Q$ can be equalized to $P_{out}$, the factor $I_C$ can be identified with $K$ – operating capital (investments) and the labour force $L$ – in the Cobb-Douglas model – corresponds to voltage $U$ taken from $R_L$ load resistor terminals (Fig. 4).

### 4.4. The correlation of economic amplifier (transistor) model with Keynes model

The investments multiplier, *m*, as defined by Keynes by the ratio [25,26]:

$$m = \frac{\Delta V}{\Delta I} \tag{35}$$

(where $\Delta V$ represents the income efficiency for $\Delta I$ investments growth by an unit), corresponds to transconductance or slope $S$ as in the case of the transistor used as model for the economic amplifier (see Eq.(20) or Eq.(22)). Similar to the electronic transistor, the output current variation $\Delta I_C$ corresponds to income efficiency, and the variation of input voltage $\Delta V_{BE}$ (or $\Delta V_{GS}$) corresponds to the investments variation increase (with an unit); therefore comparing Eq.(35) to Eq.(20) or Eq.(22), Keynes multiplier, $m$, can be identified by the economic slope, $S_{ec}$, from the economic amplifier model

$$S_{ec} = \frac{Income\ variation\ increase}{Investments\ variation,\ \Delta I} \tag{36}$$

meaning

$$S_m = m. \tag{37}$$

From the above emerges that the economic amplifier model includes several investments type models, each of them to be correlated to an electrical variable or parameter characteristic to the electronic amplifier on the basis of which the model of economic amplifier was conceived. It can be concluded that the economic amplifier (with transistors) model represents a more general model than the others, which explain only a side or an aspect of the economic concept that is analyzed through such models.

## 5. Economic development modeling using the economic amplifier model

The economic amplifier model can be very well applied for modeling the economic development based on the initial investments both at the micro and the macroeconomic levels [31]. So, for the bipolar transistor (CE configuration), the output current $I_C$ (mA) as a function of input current, $I_B$ (μA), therefore, β given by Eq.(8) has a linear dependence, resulting a constant value for β (Fig. 5*a*).

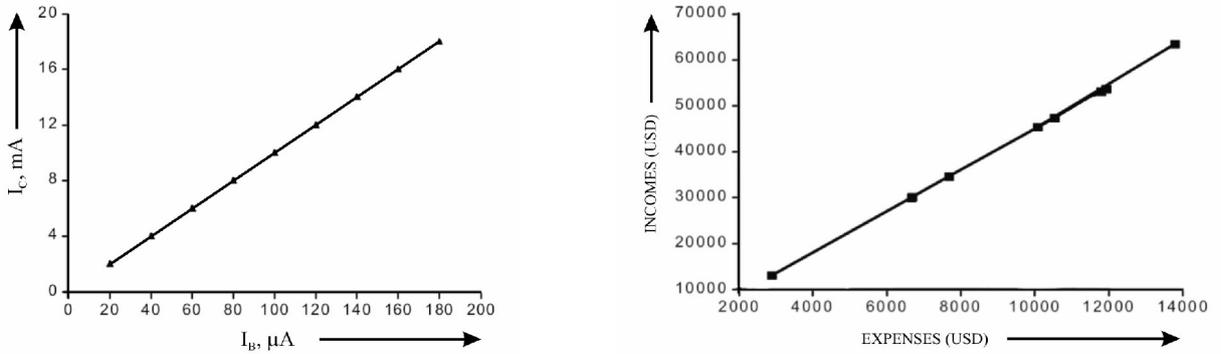

**FIG. 5** *a*) $I_C = f(I_B)$ dependence represents a straight line; *b*) Incomes representation depending of expenses ($β_{economic}$) for a small manufacture.

Indeed, the Eq.(8) can be written as

$$I_C = \beta I_B \tag{38}$$

representing a straight line (when β = const) intersecting the origin of the axes. If it is not the case and the intersection with vertical axis is $a_0$ Eq.(38) is becaming

$$I_C = a_0 + \beta I_B \tag{39}$$

also representing a straight line.

The Eq.(39) is a simple regression one exactly as the equation obtained for the one factor econometric model [27÷29] for which $a_0$, and $\beta$ are the regression coefficients. Therefore the economic amplifier model is a true econometric and implicitly an econophysics one.

Under normal conditions and economic policies by applying the economic amplifier model for the investments one may see that the incomes as a function of inputs (investments, materials etc.) are also a linear or near linear function (Figs. 5*b* and 6) justifying the econophysics model we have proposed.

An example of applying of the economic amplifier model on manufacture level (microeconomic level) is given in figure 5*b*, where the total incomes as a function of initial and annual investments and expenses are represented. As if can be seen from figure 5*b* and table 1, the ratio between total incomes and expenses (including investments), meaning $\beta_{economic}$, has a monotonous rising dependence, near a straight line.

The economic amplifier model can be also very well applied on the macroeconomic level. A good example is given in figure 6 where the dependence of the Internal Brut Product (I.B.P.) of Romania as a function of the capital accumation between 1990-2003 years is represented. As it can be seen from figure 6, the resulting function has a near linear dependence. The medium value of 5.19 for the $\beta_{economic}$ at national level is calculed.

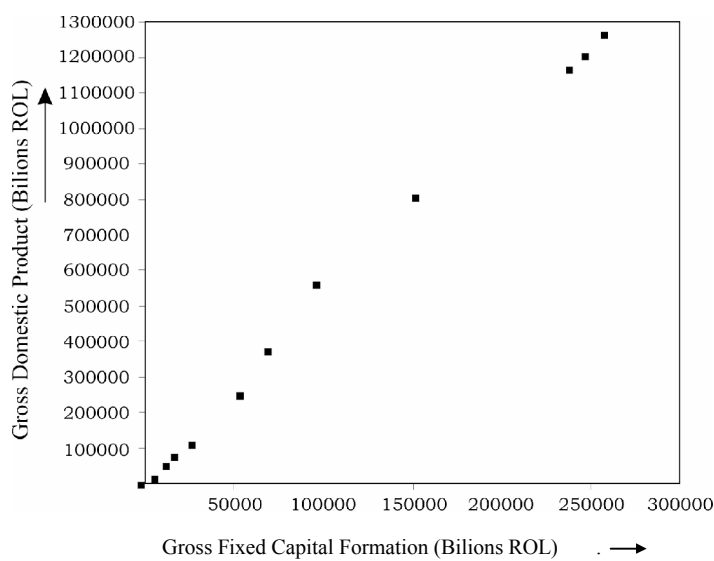

**Fig.6.** Gross Domestic Product of Romania vs. Gross Capital Formation for 1990-2004 period.

Apart from the normal development for which Incomes = *f* (Investments) dependence is a straight line or a monotonous rising curve, there are situations in which after a monotonous rising, the descending slopes appears signaling the appearance of the disturbing elements generated in special by human factor, damaging economic policies, crises, inefficient management or even political influences (totalitarian states). In such cases, when investments and equipment lifetime (*T*) had not been consumed but the curve isn't a line or a monotonous rising curve, the parameters on the basis of the investment must be reconsidered and any negative influences, restrictive policies etc. must be eliminated, in order to put the business on the normal trend of rising.

Such considerations may seem (apperar) obvious or of good sense, but it is important to remind that all these are in concordance with economic amplifier model which we have proposed here.

## 6. Conclusions

A new econophysic model, on the basis of the amplification phenomenon from appled physics, that can be realized using transistors, integrated circuits or other electronic active devices, was proposed and analyzed.

This new model, named economic amplifier, can be used for modeling different economic structures or processes such as production or investment fields, steady capital, founds or financial-banking fields stocks etc.

The accuracy of electronic transistor and of electronic amplifier functioning, confirmed through all nowadays supertechnology, can confer a solid guaranty for the economic amplifier model also in the economic fields where such model is applied.

As shown above, the economic amplifier model can explain and confirms the justness and validity of other econometric models from the investment fields.

Unlike the static models derived from statistical physics, that can not show system evolution, the economic amplifier model is a dynamic one permitting to forecast the behavior of the economic system studied depending on physical or some economical data applied on the "amplifier" input.